\documentclass{Interspeech}

\usepackage{url}
\usepackage{hyperref}
\usepackage{xspace}
\usepackage{verbatim}
\usepackage{multirow}
\usepackage{booktabs}
\usepackage{cite}
\usepackage[table,xcdraw]{xcolor}
\usepackage{array}
\usepackage{tabularx}
\newcolumntype{Y}{>{\centering\arraybackslash}X}

\usepackage{graphicx}
\usepackage{amssymb,amsmath,bm}
\usepackage{textcomp}
\usepackage{ifthen}

\newcommand{\revised}[1]{\textcolor{black}{#1}}

\interspeechcameraready

\title{Comparative Analysis of Fast and High-Fidelity Neural Vocoders for\\ Low-Latency Streaming Synthesis in Resource-Constrained Environments}

\author[affiliation={1}]{Reo}{Yoneyama}
\author[affiliation={2}]{Masaya}{Kawamura}
\author[affiliation={2}]{Ryo}{Terashima}
\author[affiliation={1,2}]{Ryuichi}{Yamamoto}
\author[affiliation={1}]{Tomoki}{Toda}

\affiliation{}{Nagoya University}{Japan}
\affiliation{}{LY Corporation}{Japan}
\email{yoneyama.reo@g.sp.m.is.nagoya-u.ac.jp, kawamura.masaya@lycorp.co.jp, ryo.terashima@lycorp.co.jp, ryuichi.yamamoto@lycorp.co.jp, tomoki@icts.nagoya-u.ac.jp}

\usepackage{comment}

\begin{document}

\maketitle

\begin{abstract}
In real-time speech synthesis, neural vocoders often require low-latency synthesis through causal processing and streaming. However, streaming introduces inefficiencies absent in batch synthesis, such as limited parallelism, inter-frame dependency management, and parameter loading overhead. This paper proposes multi-stream Wavehax (MS-Wavehax), an efficient neural vocoder for low-latency streaming, by extending the aliasing-free neural vocoder Wavehax with multi-stream decomposition. We analyze the latency-throughput trade-off in a CPU-only environment and identify key bottlenecks in streaming neural vocoders. Our findings provide practical insights for optimizing chunk sizes and designing vocoders tailored to specific application demands and hardware constraints. Furthermore, our subjective evaluations show that MS-Wavehax delivers high speech quality under causal and non-causal conditions while being remarkably compact and easily deployable in resource-constrained environments.
\end{abstract}

\section{Introduction}

Advances in deep learning have led to remarkable improvements in speech synthesis, enabling highly natural speech generation.
One key component of this progress is neural vocoders, which synthesize audio waveforms from acoustic features.
Specifically, neural vocoders based on generative adversarial networks \cite{gan}, such as MelGAN \cite{melgan} and HiFi-GAN \cite{hifigan}, have gained popularity for their favorable balance between synthesis speed and speech quality.
These models typically employ a temporal upsampling mechanism to transform low-resolution mel-spectrograms into high-resolution waveforms.
To reduce computational costs by decreasing the number of neural network layers, iSTFTNet \cite{istftnet, istftnet2} incorporates the inverse short-time Fourier transform (iSTFT) into HiFi-GAN.
More recent approaches, such as APNet \cite{apnet} and Vocos \cite{vocos}, further enhance computational efficiency by eliminating the upsampling process.
Instead, these models directly generate complex spectrograms at the frame level, avoiding operations on long sequential features.
These developments have significantly improved batch inference efficiency, typically evaluated using real-time factors measured over more than a few seconds of waveforms.

Nevertheless, this evaluation paradigm cannot adequately assess performance in low-latency streaming for real-time applications, such as live communication systems and intelligent voice assistants.
Unlike batch inference, streaming synthesis generates audio waveforms incrementally by processing small segments while maintaining constant latency and memory usage, independent of utterance length.
However, streaming synthesis introduces inefficiencies not present in batch processing \cite{snarvc}, such as limited parallelism, inter-frame dependency management, and model parameter loading overhead.
Therefore, faster batch inference does not necessarily yield higher throughput in streaming synthesis.
Furthermore, while causal processing can further reduce latency by enabling on-the-fly waveform generation \cite{llrtvc, streaming_nonar_s2s_vc, streaming_vc}, most studies assume non-causal conditions where future context is available.
These assumptions for evaluating vocoders (non-causal batch synthesis) do not accurately reflect performance constraints in low-latency scenarios.

This paper first proposes multi-stream Vocos and Wavehax (MS-Vocos and MS-Wavehax), which extend Vocos and Wavehax \cite{wavehax} by integrating multi-stream synthesis \cite{ms-hifigan}, an effective technique for enhancing computational efficiency in neural vocoders.
Second, we conduct an in-depth analysis of the relationship between latency and throughput in a CPU-only environment, to identify key bottlenecks in streaming neural vocoders.
Our findings provide practical insights into optimizing chunk sizes and designing vocoders tailored to specific application demands and hardware constraints.
For instance, minimizing latency is crucial for a comfortable user experience in real-time voice conversion \cite{dnn-rtvc, fb-dnn-rtvc, ac-vc}, whereas slight delays are tolerable in real-time text-to-speech (TTS) \cite{p2p_itts, jp_itts, st-itts}.
Third, through speech analysis-synthesis and TTS tasks, we evaluate speech quality and show that MS-Wavehax achieves superior quality compared to high-fidelity baseline models, while attaining the highest throughput under low-latency conditions ($<$ 80 ms) and requiring only 2.4\% of HiFi-GAN V1’s model size.
Moreover, with only a single-frame lookahead, MS-Wavehax achieves nearly the same quality as its non-causal counterpart, highlighting its suitability for low-latency and resource-constrained applications.

\section{Proposed method}
\label{sec:proposed method}

This section introduces Vocos \cite{vocos} and Wavehax \cite{wavehax}, along with our proposed MS-Vocos and MS-Wavehax, which integrate multi-stream synthesis \cite{ms-hifigan}.
Additionally, we elaborate on techniques to facilitate streaming feasibility.



\subsection{Overview of Vocos and Wavehax}
\label{ssec:overview of vocos and wavehax}

Vocos repurposes the ConvNeXt \cite{convnext} model, originally developed for image processing, by converting its two-dimensional (2D) convolutions into one-dimensional (1D) counterparts for speech synthesis. 
The advanced architecture has enabled estimating complex spectrograms without upsampling, achieving high speech quality and extremely fast batch synthesis.
However, working with complex spectrograms in the time-frequency domain lacks an effective inductive bias for harmonic modeling. 
As a result, Vocos exhibits reduced robustness to unseen data, particularly variations in the fundamental frequency ($F_0$).

To address this limitation, Wavehax incorporates a harmonic prior signal, using its complex spectrogram as an initial input. It then applies ConvNeXt-based 2D convolutions to preserve the spectral structure of the harmonic spectrogram. Consequently, Wavehax can be regarded as a 2D convolutional variant of Vocos, enhanced with the harmonic prior.

\subsection{Subscale synthesis}
\label{ssec:subscale synthesis}

Subscale processing significantly improves the computational efficiency of neural vocoders.
Neural vocoders using this technique produce subscale signals and reconstruct the signal with pseudo-quadrature mirror filters (PQMF) \cite{pqmf, mb-melgan}, discrete wavelet transform (DWT) \cite{fregan, fregrad}, or trainable filters \cite{ms-hifigan}. 
In our internal experiments with four subscales, trainable filters slightly outperformed PQMF and DWT in objective metrics \cite{pesq, utmos} across tested vocoders \cite{hifigan, istftnet, vocos, wavehax}.
Based on this finding, we incorporate subscale synthesis with trainable filters into Vocos and Wavehax as follows.

\noindent \textbf{MS-Vocos}:
MS-Vocos largely retains the standard Vocos architecture \cite{vocos}.
Following \cite{ms-istft-hifigan}, its final layer is split along the channel axis into four sub-spectrograms, each representing a stream’s complex spectrogram.
Consequently, this layer produces 968 output channels, which comprise log-magnitude and implicitly encoded phase information for each subscale signal.
MS-Vocos then applies iSTFT with a frame length of 240 (40 ms) and a frame shift of 60 (10 ms), one-fourth of the standard Vocos settings (960 and 240, respectively).
Finally, the subscale waveforms reconstructed by iSTFT are combined via a synthesis filter to produce the final waveform.

\begin{figure}[tb]
    \begin{center}
    \includegraphics[width=0.98\columnwidth]{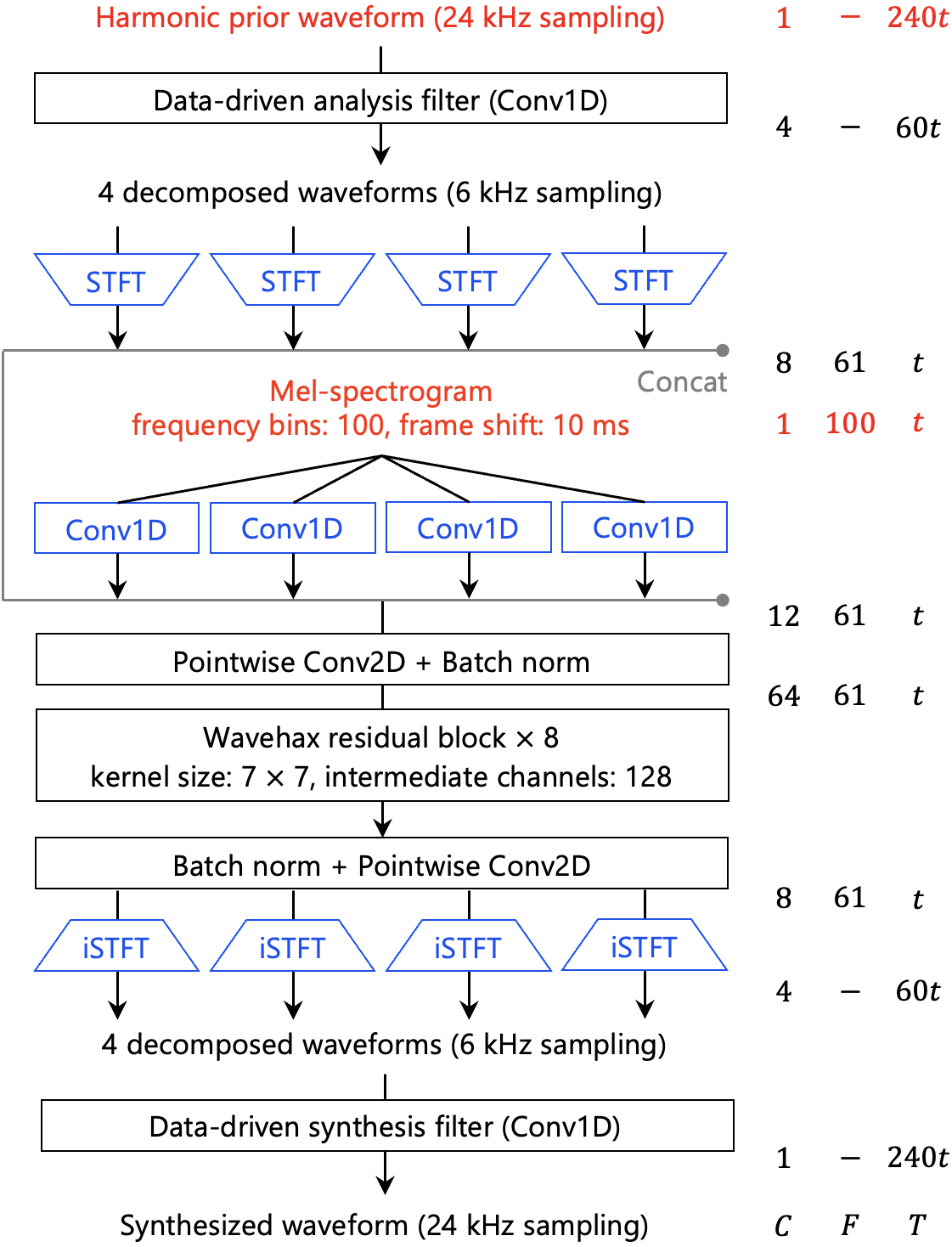}
    \end{center}
    \vspace{-4.5mm}
    \caption{\small An overview of multi-stream Wavehax. The harmonic prior waveform is generated from the input fundamental frequencies via signal processing. The lengths of the analysis and synthesis filters are both 63. The frame length and frame shift for STFT/iSTFT are 20 ms and 10 ms, respectively. The numbers on the right represent the tensor shapes at each processing stage, where $C$, $F$, $T$, and $t$ denote the number of channels, frequency bins, sequential length, and input length, respectively.}
    \label{fig:ms-wavehax}
    \vspace{-3.5mm}
\end{figure}

\noindent \textbf{MS-Wavehax}:
MS-Wavehax differs from MS-Vocos and prior vocoders \cite{ms-hifigan, ms-istft-hifigan} by employing an analysis filter to decompose the input signal (Fig.~\ref{fig:ms-wavehax}).
This process produces four subscale waveforms, each undergoing STFT to yield a corresponding two-channel complex spectrogram. 
These spectrograms are then concatenated with the projected mel-spectrogram features, resulting in a 12-channel representation.
These spectrograms are then concatenated with the projected mel-spectrogram features, resulting in a 12-channel representation that is processed by a series of convolutional layers and residual blocks.
MS-Wavehax employs 64 hidden channels, twice that of standard Wavehax, while its frequency dimension is reduced to one-fourth. 
Consequently, it requires approximately half the computational cost of Wavehax.
The final pointwise Conv2D layer outputs eight spectrograms (real and imaginary components of the subscale signals), which are then converted back to the time domain via iSTFT.
Finally, a synthesis filter merges the subscale signals into the final waveform.


\subsection{Streaming feasibility}
\label{ssec:streaming feasibility}

For efficient streaming synthesis, we introduce two key techniques.
First, we apply caching mechanisms \cite{caching_mechanism} to all convolutional, STFT, and iSTFT layers. 
The caching mechanisms use ring buffers to retain and reuse intermediate inputs or outputs from previous segments, thereby eliminating redundant computations.
\revised{Second, we replace the layer normalization (LN) \cite{layer_norm} layers with batch normalization (BN) \cite{batch_norm} layers, as Wavehax's LN implementation computes statistics over the entire sequence, making it unsuitable for streaming.
In contrast, BN does not require computing statistics during inference, making it more appropriate for streaming settings.
We apply the same modification to Vocos.
Interestingly, this change resulted in improved objective metrics \cite{pesq, utmos} in our internal experiments.}

\section{Analysis of latency vs. throughput}
\label{sec:analysis of latency vs. throughput}

We analyze the relationship between latency and throughput via block streaming synthesis using several neural vocoders.

\subsection{Model details}
\label{ssec:model details}

We compared four neural vocoders, HiFi-GAN (V1) \cite{hifigan}, iSTFTNet (V1) \cite{istftnet}, Vocos \cite{vocos}, and Wavehax \cite{wavehax}, along with their multi-stream variants, each configured with four streams.
All models retained their original configurations unless otherwise specified.
The upsampling factors were 8, 5, 3, and 2 for HiFi-GAN; 5, 3, 2, and 2 for MS-HiFi-GAN \cite{ms-hifigan}; 10 and 6 for iSTFTNet; and 5 and 3 for MS-iSTFTNet \cite{ms-istft-hifigan}.
Both iSTFTNet and MS-iSTFTNet used a DFT size of 16 and a hop size of 4.
All models assumed a 100-dimensional mel-spectrogram and a 1-dimensional $F_0$ input.
Wavehax and MS-Wavehax utilized $F_0$ for generating prior signals, whereas other models concatenate it with the mel-spectrogram, resulting in a 101-dimensional input feature.
The sampling rate and frame shift are fixed at 24 kHz and 10 ms, respectively.
Each model was adapted for streaming, following the techniques described in Section~\ref{ssec:streaming feasibility}.

\begin{table}[tb]
\renewcommand{\arraystretch}{1.25}
\caption{\small Parameter counts (Params, millions) and multiply-accumulate operations (MACs, billions) for generating one second of audio. To support ONNX conversion, STFT/iSTFT layers (torch.stft and torch.istft) were replaced with convolution-based implementations, resulting in higher MACs.}
\vspace{-2.5mm}
\small
\centering
\scalebox{0.88}{
\begin{tabularx}{\columnwidth}{p{2cm}>{\centering\arraybackslash}YYY}
    \cline{1-4}
    & \multirow{2}{*}{Params $\downarrow$} & \multicolumn{2}{c}{MACs} \\
    &  & PyTorch & ONNX \\
    \cline{1-4}
    HiFi-GAN    & 13.82 & 28.02 & 28.02 \\
    MS-HiFi-GAN & 10.71 & 12.89 & 12.89 \\
    iSTFTNet    & 13.73 & 25.85 & 25.83 \\
    MS-iSTFTNet & 12.30 & 8.370 & 8.391 \\
    Vocos       & 13.50 & 1.348 & 46.96 \\
    MS-Vocos    & 13.50 & 1.356 & 4.215 \\
    Wavehax     & 0.622 & 1.787 & 13.03 \\
    MS-Wavehax  & 0.332 & 1.576 & 2.291 \\
    \cline{1-4}
    \end{tabularx}
}
\label{table:params_macs}
\renewcommand{\arraystretch}{1.0}
\vspace{-2.5mm}
\end{table}


\subsection{Results and discussion}
\label{ssec:results and discussion}

\begin{figure}[tb]
    \vspace{-3mm}
    \begin{center}
    \includegraphics[width=\columnwidth]{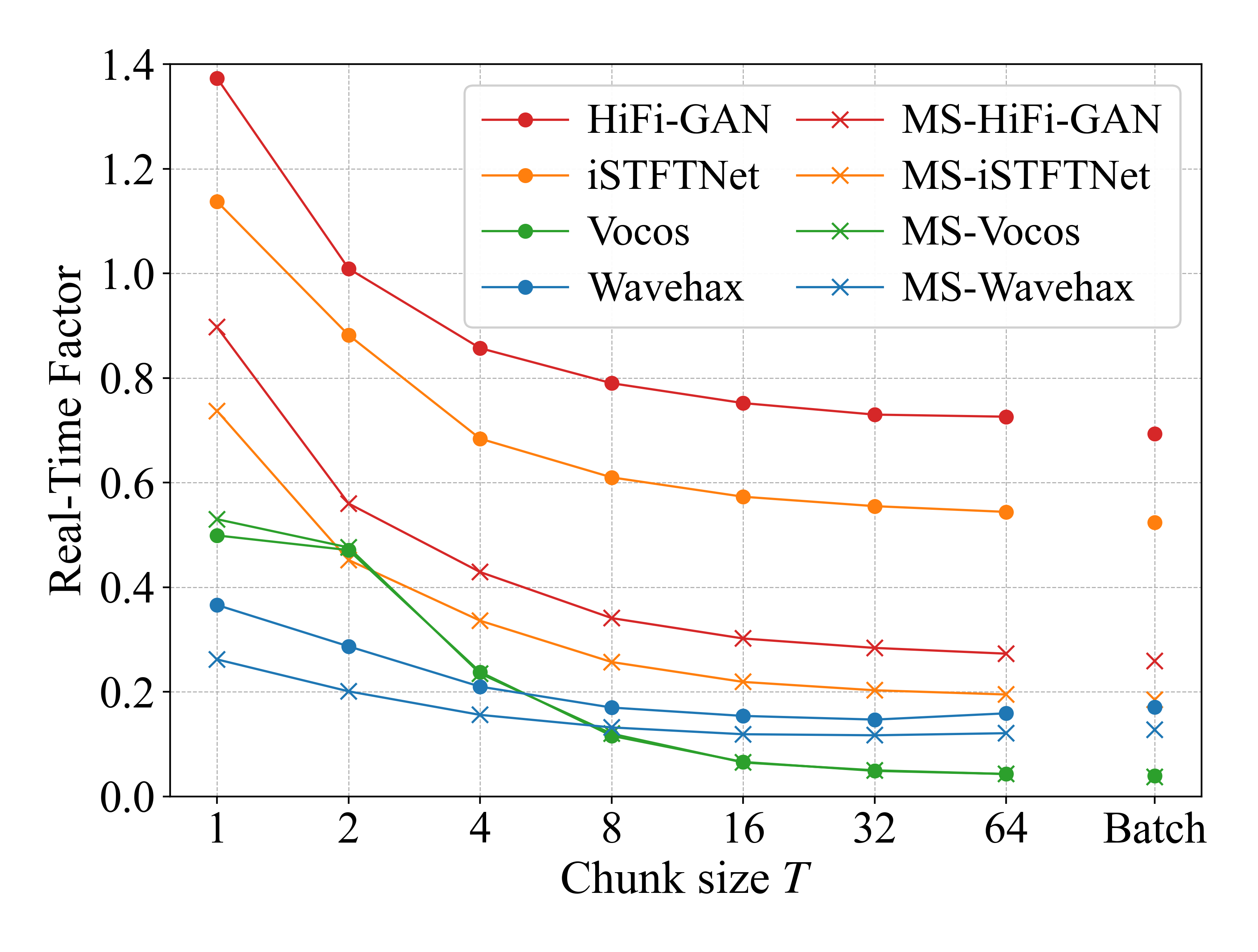}
    \end{center}
    \vspace{-8mm}
    \caption{\small 
    Real-time factors for batch and streaming synthesis, measured using ONNX Runtime on a single thread of an AMD EPYC 7302 processor (maximum clock speed: 3 GHz). 
    The results were averaged over 30 runs, each lasting 10 seconds, following 5 initial warm-up iterations.
    The chunk size $T$ is the number of time frames generated per synthesis step.
    For example, a chunk size of 8 introduces an additional latency of 70 ms relative to a chunk size of 1, as the vocoder must wait for input features of all 8 frames before starting synthesis.}
    \label{fig:rtf}
    \vspace{-3mm}
\end{figure}

\begin{figure}[tb]
    \begin{center}
    \vspace{-2.7mm}
    \includegraphics[width=0.98\columnwidth]{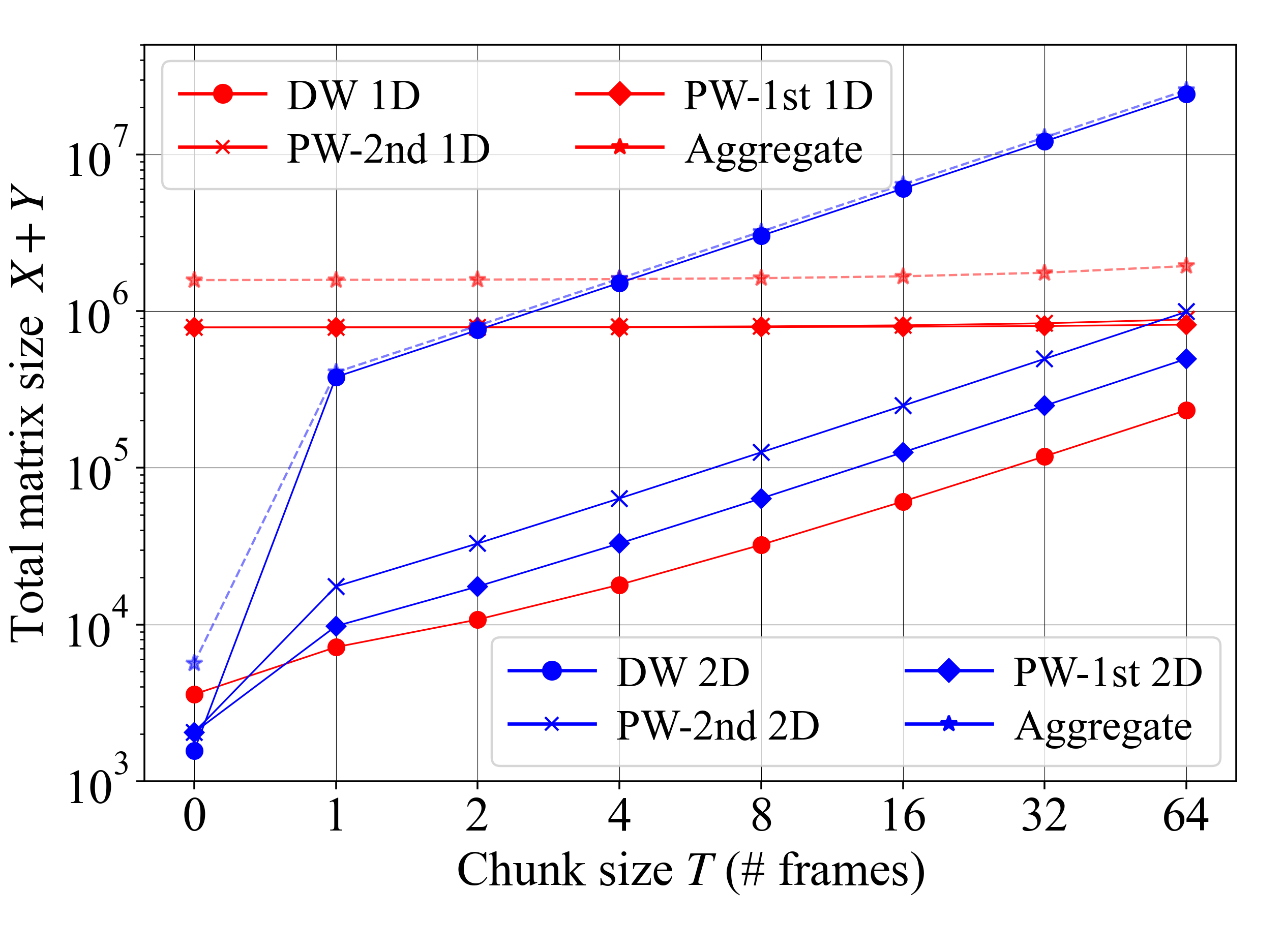}
    \end{center}
    \vspace{-8.3mm}
    \caption{\small The theoretical total matrix size $X + Y$ as a function of the chunk size $T$, where $X$ and $Y$ are defined in Table~\ref{table:im2col} as the input and filter matrix sizes, respectively. \revised{“Aggregate”} represents the total matrix size, computed as \revised{DW + PW-1st + PW-2nd}. The y-intercept corresponds to the filter matrix size.}
    \label{fig:matrix_size}
    \vspace{-4mm}
\end{figure}

Figure~\ref{fig:rtf} illustrates how increasing chunk size, which leads to higher latency, affects the real-time factor (RTF) in streaming synthesis.
Increasing the chunk size tends to accelerate synthesis by reducing overhead from cache management and parameter loading.
For smaller chunk sizes, MS-Wavehax achieves the lowest RTF, followed by Wavehax, due to fewer parameters and lower multiply-accumulate operations (MACs) (see Table~\ref{table:params_macs}).
Conversely, at larger chunk sizes, Vocos achieves the highest throughput.
To explore this RTF reversal, we first examine the matrix sizes produced by the im2col operation, which reformulates convolution as matrix multiplication.
Table~\ref{table:im2col} compares the matrix sizes and MACs in the residual blocks of Vocos and Wavehax
\footnote{Although our analysis focuses on a particular scenario where backend-level optimizations depend on $im2col$, not all optimizations use this technique. Nonetheless, because depthwise and pointwise convolutions involve fewer operations than standard convolutions, their data transfer behavior is expected to follow similar trends across various implementations (e.g., direct convolution).}.
Drawing on Table~\ref{table:im2col}, Fig.~\ref{fig:matrix_size} depicts how the matrix sizes vary with the chunk size $T$.
In 2D convolutions, the matrix size of the input tensor scales with $H_o$ (the frequency dimension) and grows faster than in 1D convolutions as $T$ increases.
Larger $im2col$ matrices require more frequent access to main memory, which is less efficient than accessing cache memory.
In contrast, the filter matrices of Vocos’s 1D convolutions are larger than those of Wavehax’s 2D convolutions; yet remain constant regardless of $T$.
These kernels are repeatedly reused throughout the convolution process, increasing the cache hit rate.
Hence, as the chunk size grows, data-transfer overhead becomes the primary bottleneck in 2D convolutions.
Notably, the relationship between latency and streaming throughput is influenced by hardware configurations, including processor speed, cache hierarchy, and memory bandwidth\footnote{This experiment used a processor with a three-level cache hierarchy: each core had a 64 KiB L1 cache (32 KiB for data and instructions), a 512 KiB L2 cache, and a shared 16 MiB L3 cache.}.
Slower processors, particularly those lacking SIMD (single instruction, multiple data) instructions or specialized accelerators, are more susceptible to computational costs (i,e., MACs).
A limited cache capacity raises the frequency of cache misses, leading to higher data and parameter transfer overhead and greater dependency on memory bandwidth.

\begin{table*}[tb]
\renewcommand{\arraystretch}{1.3}
\vspace{-1mm}
\caption{\small 
Comparison of matrix representations and computational complexity in depthwise (DW) and pointwise (PW) convolutions, which constitute a residual block in Vocos \cite{vocos} and Wavehax \cite{wavehax}, with a chunk size $T$.
Convolution of an input tensor of shape $(B, C_i, H_i, W_i)$ with a filter of shape $(C_i, C_o, H_f, W_f)$ can be efficiently implemented using the $im2col$ operation and General Matrix Multiplication (GeMM).
Specifically, the convolution is performed via matrix multiplication: $(B H_o W_o, C_i H_f W_f) \times (C_i H_f W_f, C_o) \rightarrow (B H_o W_o, C_o)$.
Here, $B$ is the batch size, $C_i$ and $C_o$ are the input and output channels, and $H_i$, $W_i$, $H_o$, $W_o$, $H_f$, and $W_f$ denote the height and width of the input, output, and filter tensors.
Since depthwise convolution processes each input channel independently, it can be reformulated as a standard convolution by treating the channel dimension as an additional batch dimension.}
\vspace{-2mm}
\centering
\small
\scalebox{0.88}{
\begin{tabular}{lcccccc}
    \cline{1-7}
    & Input shape & Column shape & Filter shape & Column size $X$ & Filter size $Y$ & MACs \\
    & $(B, C_i, H_i, W_i)$ & $(B, H_o, W_o, C_i, H_f, W_f)$ & $(C_i, H_f, W_f, C_o)$ & $B H_o W_o C_i H_f W_f$ & $C_i H_f W_f C_o$ & $B H_o W_o C_i H_f W_f C_o$ \\
    \cline{1-7}
    DW 1D & $(1, 512, 1, T)$ & $(512, 1, T, 1, 1, 7)$ & $\revised{512 \, \times} \,(1, 1, 7, 1)$ & $3584 \times T$ & \revised{$3584$} & $3584 \times T$ \\
    PW-1st 1D & $(1, 512, 1, T)$ & $(1, 1, T, 512, 1, 1)$ & $(512, 1, 1, 1536)$ & $512 \times T$ & $786432$ & $786432 \times T$ \\
    PW-2nd 1D & $(1, 1536, 1, T)$ & $(1, 1, T, 1536, 1, 1)$ & $(1536, 1, 1, 512)$ & $1536 \times T$ & $786432$ & $786432 \times T$ \\
    \cline{1-7}
    DW 2D & $(1, 32, 241, T)$ & $(32, 241, T, 1, 7, 7)$ & $\revised{32 \, \times} \,(1, 7, 7, 1)$ & $377888 \times T$ & \revised{$1568$} & $377888 \times T$ \\
    PW-1st 2D & $(1, 32, 241, T)$ & $(1, 241, T, 32, 1, 1)$ & $(32, 1, 1, 64)$ & $7712 \times T$ & $2048$ & $493568 \times T$ \\
    PW-2nd 2D & $(1, 64, 241, T)$ & $(1, 241, T, 64, 1, 1)$ & $(64, 1, 1, 32)$ & $15424 \times T$ & $2048$ & $493568 \times T$ \\
    \cline{1-7}
    \end{tabular}
}
\label{table:im2col}
\vspace{-2mm}
\renewcommand{\arraystretch}{1.0}
\end{table*}

\section{Speech quality evaluation}
\label{sec:speech quality evaluation}

In the experiments in Section~\ref{sec:analysis of latency vs. throughput}, MS-Wavehax achieved the highest throughput under low-latency conditions.
Next, we evaluate its speech quality under causal and non-causal conditions, compared to the vocoders described in Section~\ref{ssec:model details}.

\subsection{Preparation}
\label{ssec:preparation}

\noindent \textbf{Analysis-synthesis (A/S) setup}:
We used the JVS corpus \cite{jvs}, containing 30.2 hours of speech data from 100 Japanese speakers. 
Eight speakers (No. 1-8) were allocated for validation (800 utterances) and evaluation (400 utterances), while the rest (No. 9-100) were used for training.
The vocoders were conditioned on 100-bin mel-spectrograms and $F_0$ values.
Mel-spectrograms were computed via STFT with a 1024-point DFT, and a frequency range of 0–8 kHz.
$F_0$ values were extracted with the Harvest algorithm \cite{harvest}.
(MS-)HiFi-GAN and (MS-)iSTFTNet were trained following the optimizer settings in \cite{bigvgan}, while (MS-)Vocos and (MS-)Wavehax followed those in \cite{vocos}.
Gradient clipping was applied with a threshold of 10.
All models were trained for 1M steps using the UnivNet \cite{univnet} discriminator.
We set the batch size to 16 and the batch length to 32 frames.

\noindent \textbf{TTS setup}:
We used the JSUT corpus \cite{jsut}, comprising 10.3 hours of speech data spoken by a Japanese female speaker.
The vocoders were conditioned on mel-spectrograms and $F_0$ values, both predicted from text inputs.
We generated mel-spectrograms using the text-to-mel predictor in Matcha-TTS \cite{matchatts} (Mel-model), which combines a Transformer encoder \cite{transformer} with a decoder based on optimal-transport conditional flow matching \cite{flow_matching}.
We used the official implementation \cite{matchatts_code} and set the mel-spectrogram hyperparameters to match those of the A/S task.
We developed an $F_0$ prediction model ($F_0$-model) based on the ConvNeXt architecture \cite{convnext} to estimate voiced/unvoiced probabilities and log-scale $F_0$ values for each time frame, conditioned on input mel-spectrograms.
The Mel-model was trained for 3K epochs on the BASIC5000 subset (approximately 6.6 hours), where phoneme labels with accent information are available \cite{jsut_label}.
From BASIC5000, 180 utterances were used for evaluation and 20 for validation.
The $F_0$-model and vocoders were trained on the entire dataset (about 10.1 hours), excluding the evaluation and validation sets.
We trained the $F_0$-model for 500K steps with mel-spectrogram and $F_0$ sequence pairs extracted from natural speech through supervised learning.
The vocoders were fine-tuned for another 100K steps using estimated acoustic features.

\subsection{Evaluation results}
\label{ssec:evaluation results}

\begin{figure}[tb]
    \vspace{-1mm}
    \includegraphics[width=\columnwidth]{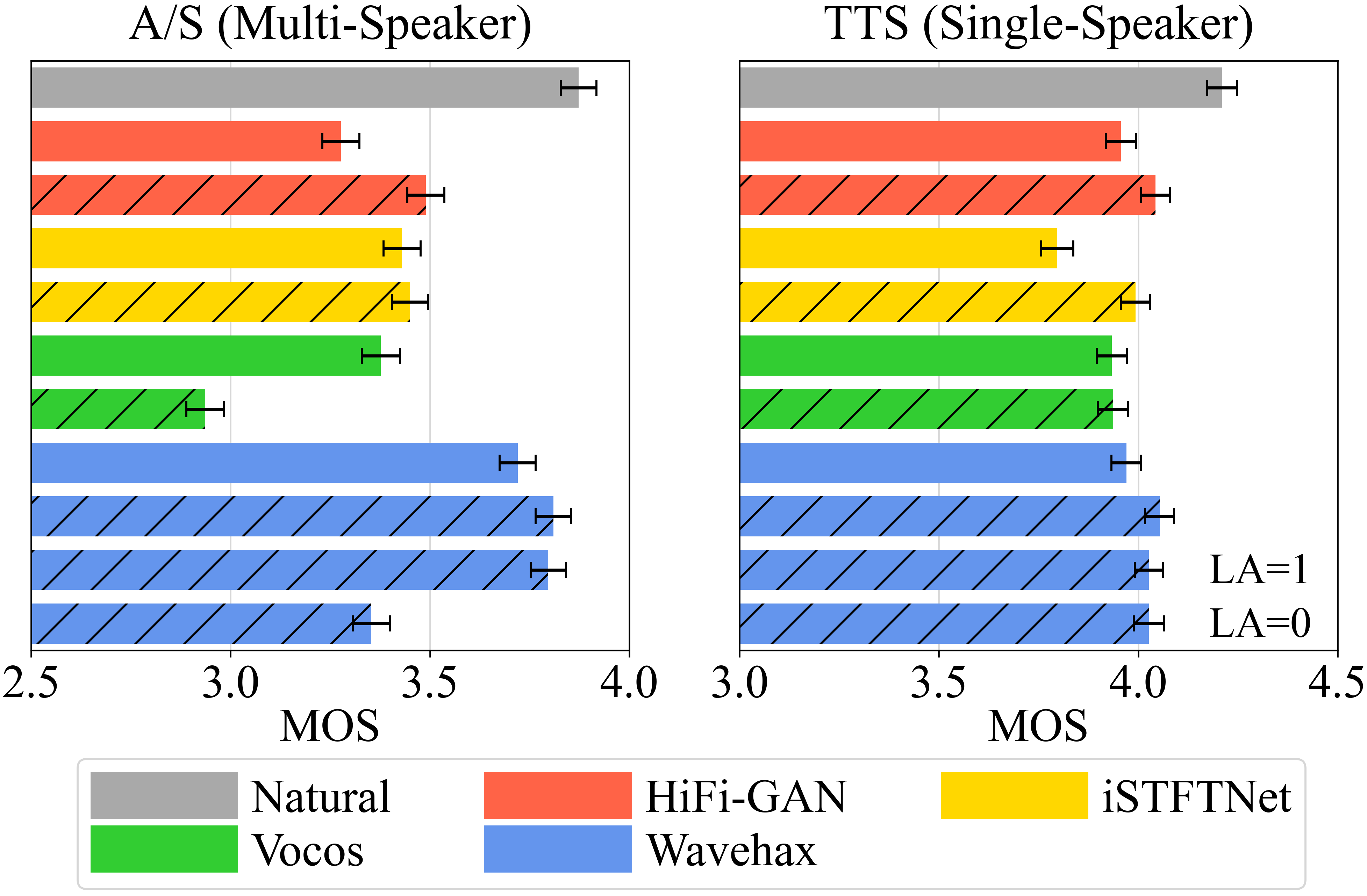}
    \vspace{-5mm}
    \caption{\small Five-point mean opinion scores (MOS) on speech quality with 95\% confidence intervals. Thirty native Japanese speakers evaluated 10 samples per model for analysis-synthesis (A/S) and text-to-speech (TTS). Slashes in the bars indicate multi-stream models. The bottom two bars represent causal models with lookahead (LA) of 1 and 0.}
    \label{fig:mos}
    \vspace{-4mm}
\end{figure}

Figure~\ref{fig:mos} presents the MOS on speech quality in the A/S and TTS tasks.
We evaluated non-causal HiFi-GAN, iSTFTNet, Vocos, and Wavehax, along with their multi-stream variants, as well as causal MS-Wavehax with or without one-frame lookahead (LA) \cite{streaming_nonar_s2s_vc, lah4itts, llrtvc}.
As illustrated, non-causal MS-Wavehax and causal MS-Wavehax with LA achieved the highest MOS in both tasks.
For TTS, differences among all methods, including natural speech, were smaller, likely because the single-speaker setting simplifies TTS compared to A/S, resulting in more uniform quality.
Additionally, the performance gap between MS-Wavehax with LA=1 and LA=0 disappeared in TTS.
Possible reasons include: (1) stricter causal constraints have a greater impact in multi-speaker scenarios, (2) MS-Wavehax’s small parameter count makes it relatively sensitive to input acoustic features, and (3) causality served as an effective regularization, improving generalization in TTS.

Regarding the multi-stream variants, MS-HiFi-GAN and MS-iSTFTNet outperform their non-MS counterparts, consistent with previous findings \cite{ms-hifigan, ms-istft-hifigan}.
The improvements of MS-Wavehax over Wavehax can be attributed to the spatial shift-invariance of 2D convolutions.
While full-band spectrograms do not exhibit uniform spatial structures across frequencies, Wavehax applies the same convolutional kernels uniformly across spatial dimensions.
In contrast, MS-Wavehax partitions the spectrogram into smaller sub-spectrograms with an arbitrary decomposition, where spectral structures tend to be more locally consistent.
This facilitates efficient use of convolutional kernels, allowing them to better capture dependencies within and across sub-spectrograms, thereby improving speech quality.
Conversely, MS-Vocos underperforms compared to Vocos.
We speculate that estimating inter-stream dependencies is particularly challenging for MS-Vocos, due to its constrained latent representation, which has fewer channels than the output complex spectrum.
Although increasing latent channels can improve expressiveness, it also raises computational costs, offsetting the efficiency benefits of subscale processing.
Therefore, we retain the original 512 channels, to preserve efficiency.

\section{Conclusion}
\label{sec:conclusion}

This study examined neural vocoders for streaming applications.
Our analysis revealed that streaming throughput depends on overhead from data and parameter loading as well as computational complexity.
This overhead varies with the streaming unit sizes and the choice of 1D and 2D convolutions.
These findings provide insights into designing streaming neural vocoders for real-world deployment.
Our experiments also showed that the proposed MS-Wavehax demonstrated high audio quality and throughput under low-latency conditions.

\section{Acknowledgements}
This work was supported in part by JST AIP Acceleration
Research JPMJCR25U5, Japan, and in part by the Japan Society for the Promotion of Science (JSPS) Grants-in-Aid for Scientific Research (KAKENHI) under Grant 24KJ1236.

\bibliographystyle{IEEEtran}
\bibliography{mybib}

\end{document}